A Master Thesis

Submitted to the

Faculty

of

American Public University

System by

Alan T. Dugger

In Partial Fulfillment of the

Requirements for the Degree

of

Master of Science in Space Studies

June 2024

American Military

University Charles Town,

WV

The author hereby grants the American Public University System the right to display these contents for educational purposes.

The author assumes total responsibility for meeting the requirements set by United States copyright law for the inclusion of any materials that are not the author's creation or in the public domain.

© Copyright 2024 by Alan Dugger

All rights reserved.

# DEDICATION

I dedicate this thesis to my wife, Raelene and our beautiful daughters, Emmalene, and Madeleine. Raelene, without your encouragement and support, I could never have gotten this far or been half as successful. Emmie and Maddie, remember it is not just the world, but the entire universe that is within your reach. There is nothing you cannot achieve.

# ACKNOWLEDGEMENTS

"I have loved the stars too fondly to be fearful of the night."

- Sarah Williams from 'The Old Astronomer to His Pupil'

I wish to thank all of the professors and fellow students that have guided me through this degree. Without your instruction, help, and insight, I would not have gained so much information and knowledge during this program. I especially want to thank the members and faculty of the university's Supernovae Research Group and the Analogue Astronaut Research Group, both of which I can proudly say I am a legacy member. The discussions and work we completed together made me confident in my own abilities but challenged me to question my own ideas and ensure they were grounded in science.

This degree has been one of the most rewarding and humbling experiences of my life.

# ABSTRACT

This thesis conducts a systematic review of the applications of Lagrange points within the solar system, utilizing Systems Theory to frame these applications in terms of their interdependencies and potential for integration into broader space mission architectures. By analyzing various applications across domains such as astronomical research, space exploration, space resource utilizations, national defense, and space communication, this study identifies key areas where Lagrange points offer significant advantages. The research employs a qualitative analysis of existing literature combined with theoretical modeling to demonstrate how these points can be optimally utilized in future space missions. The finding suggests that a Systems Theory approach not only clarifies the roles and benefits of Lagrange points in space mission design but also reveals new pathways for enhancing mission efficiency and effectiveness. This thesis underscores the importance of a holistic view in the strategic planning of space missions and provides a foundational approach for integrating Lagrange points into future exploratory and operational frameworks.

# TABLE OF CONTENTS

# LIST OF FIGURES

# CHAPTER ONE: INTRODUCTION

"Look up at the stars, and not down at your feet. Try to make sense of what you see and wonder about what makes the universe exist. Be curious."

- Stephen Hawking

**Chapter Introduction**

The exploration and utilization of space have become central themes in advancing human knowledge and capabilities beyond Earth. Among the many strategic points of interest in space, Lagrange points – locations in space where the gravitational forces of a two-body system, like Earth and the Sun, produce enhanced regions of attraction and repulsion – offer unique opportunities (Parker & Anderson, 2014). These points allow for minimal fuel use in station-keeping, making them ideal for various space mission applications ranging from astronomical observatories to defense platforms. The exploitation and utilization of Lagrange points represents an opportunity in space exploration and technology, presenting distinct advantages for the missions stated above and those yet undiscovered.

**Statement of the Problem**

These points have been the subject of extensive theoretical research and practical application for decades (Steg & De Vries, 1966). Yet, despite their potential, the systematic study of Lagrange point applications within a cohesive framework has been limited. Previous research has often segmented these applications by specific domains, which this thesis has categorized as Areas of Focus, astronomical research, space exploration, resource utilization, national defense, and space communication. The full benefit of Lagrange points as segments within a larger space system may be lost without a holistic approach that aims to understand the interdependencies and collective impact each domain has on space mission design.

This thesis aims to systematically examine the breadth and depth of Lagrange point applications, delineating the theoretical underpinnings that guide their use and the myriad ways in which they have been leveraged for space missions. By doing so, it seeks to propose a review of the current state of knowledge in this area. Further, by applying systems theory to explore the interdependent nature of space mission designs that utilize Lagrange points, this thesis aims to bridge the gap between current applications and potential future uses when Lagrange points are considered components of a larger and more dynamic space mission concept.

**Background and Need**

The history of our exploration of space has seen thousands of satellites placed into Earth orbit, and robotic landers and rovers explore nearby bodies like planets, moons, and asteroids. The International Space Station (ISS), currently in low Earth orbit, may constitute the greatest feet of astronomical engineering, perhaps second only to the Apollo missions that saw human beings land and walk on the Moon and then returned them safely to Earth.

Yet today, more entities than ever are setting their sights on space; reaching further and deeper into the cosmos than we have ever gone before (Iliopoulos & Esteban, 2020). So many of the missions that have constituted the history of human exploration have often been narrowly focused where components of the mission were designed in isolation without considering their interdependencies with other missions and instead focused solely on their individual objectives. In these designs, satellites, platforms, rovers, and even spacecraft become debris and relics of past ideas. As government and private sector organizations pursue more dynamic space initiatives, such as NASA's Lunar Gateway and missions aimed at human settlement on other planets, it becomes crucial to view Lagrange points and other mission components as integral to the broader strategy for solar system exploration and possible colonization.

**Purpose of the study**

The purpose of this study was to systematically review the applications of Lagrange points within the solar system, specifically focusing on their role in astronomical research, space exploration, space resource utilization, national defense, and space communication. By applying Systems Theory, this research aims to demonstrate the interdependent nature of space missions that utilize these points. It seeks to identify and analyze potential redundancies and efficiencies, thereby proposing optimized strategies for future space missions and infrastructure development. This study not only fills gaps in current literature by providing a holistic view of Lagrange point applications but also offers strategic insights beneficial for policymakers, engineers, and scientists involved in the planning and execution of complex space missions.

The structure of this thesis is as follows: Chapter two reviews the literature on the applications of Lagrange points, identifying current issues and the potential for future developments. Chapter three outlines the theoretical framework of systems theory as applied to space missions. Chapter four details the methodology employed to analyze the interdependencies of Lagrange point applications. Chapter five presents the findings and discusses their implications for future space mission planning and executions. Finally, chapter six concludes with a summary of the research, its limitations, and suggestions for further study.

**Research Question**

This thesis seeks to answer a critical question formulated in the hypotheses presented in chapter three: Do space missions exhibit interdependent relationships that influence their success in achieving their objectives? Furthermore, it explores whether applying Systems Theory to conduct an integrated analysis of missions utilizing Lagrange points could reveal more efficient approaches and broader objectives.

**Significance to the Field**

A systematic review of Lagrange point applications could introduce a novel framework for analyzing and optimizing complex interdependencies in space mission design, leading to a more holistic and adaptable approach for space missions that are increasingly involving multiple stakeholders and complex technologies. Further, demonstrating how Systems Theory can uncover efficiencies not evident when missions are planned in isolation and can inform and assist policymakers, engineers, and scientists in the strategic planning and decision making for future investments in space infrastructure, mission scope, and objectives. Finally, as are the hopes of all researchers, this thesis hopes to push the boundaries of current scientific knowledge and technological concepts in this field and provide a deeper understanding of how space missions can be designed to be more synergistic.

**Terms and Definitions**

This thesis presents many terms that may be common or easily understood to the reader. Obscure or infrequently used terms are defined in their context within the thesis itself. Acronyms and abbreviations are identified in full before later using their shortened forms.

This thesis does present a taxonomy for classifying the Lagrange points in the solar system, a concept that does not presently have a universally accepted format in literature. The Lagrange points in the solar system will be identified by the two bodies that make up what this research presents as a "Lagrangian system" or the family of Lagrange points that exist due to the influence of the two bodies whose centrifugal and gravitational forces create them. The larger of the two bodies will be listed first and the small body second with the Lagrange point being discussed listed subsequently with an "L" and an Arabic numerical identifier to denote its position. For example, to discuss the first Lagrange point of the Sun and Earth Lagrangian system, this paper uses "Sun-Earth L1." Here, the Sun is the larger of the two bodies, the Earth is the smaller, and the first Lagrange point is identified by "L1." All five of the Lagrange points that exist due to the influence that the Sun and Earth exert on each other (Sun-Earth L1, Sun-Earth L2, Sun-Earth L3, Sun-Earth L4, and Sun-Earth L5) are collected within the term "the Sun-Earth Lagrangian system."

Within each Lagrangian system, the Lagrange points are further categorized by the mathematical formulas used to determine their positions. When viewed from above, the first three Lagrange points in any Lagrangian system lie in a straight line that intersects the centers of the large and small bodies of the system, thus these three points are referred to as the "collinear points." The fourth and fifth points that lie along the orbital path of the smaller body around the larger body are presented at 60 degrees ahead of and behind the smaller body as it orbits 360 degrees around the larger body. As such, the fourth and fifth points create an equilateral triangle between themselves and the centers of the two bodies that constitute their Lagrangian system. Thus, these two points are referred to as the "equilateral points."

**Limitations**

This research will have several limitations, that while significant, will not detract from the information and knowledge presented. First, all sources were reviewed for their academic integrity and the credibility of the authors of those sources were considered before being included in the review. Only sources grounded in scientifically proven concepts or ideas were included in the literature review or in subsequent chapters. Only unclassified and open-source materials were used in the study, and it is possible that confidential research may be present or on-going at this time that could either add to the knowledge gained in the reading of this thesis or even refute many of the ideas presented. Secondly, it is not without acknowledgement from the author that there may be ideas related to Lagrange point applications in current mission design and proposals that were missed due to the proliferation of online publications over the past few decades. Despite the use of new academically based artificial intelligence research programs, it is possible that additional sources that could reinforce some segments of the research were overlooked.

**Ethical Considerations**

No human or animal participants or interviewees were conducted in the course of this research. Any ethical considerations were aimed at the integrity of the research reviewed and the authors accurate interpretation of the concepts and ideas presented. While artificial intelligence research programs and large language models were used in the location of relevant materials and the refinement of the language and presentation of the information to ensure clarity and enhance comprehension, it is important to note that the concepts, ideas, and overall content are original and have been developed independently by the author.

# CHAPTER TWO: LITERATURE REVIEW

**Chapter Introduction**

Given the rapid advancements in space technology and the increasing interest in sustainable and efficient space exploration (Iliopoulos & Esteban, 2020), understanding the applications of Lagrange points is more pertinent than ever. This review is contextualized within a broader thesis that investigates both the theoretical frameworks and practical implementations of Lagrange point applications, with a special emphasis on how these theoretical concepts have been translated into real-world projects, such as the James Webb Space Telescope currently residing at the Sun-Earth second Lagrange point, or L2 (Menzel et al., 2023). Through a thorough examination of existing literature, this review will lay the groundwork for a comprehensive analysis of Lagrange point applications, setting the stage for subsequent discussions on their potential, challenges, and future directions.

This review is organized into distinct sections, each focusing on a critical aspect of Lagrange points and their multifaceted applications: astronomical research applications, space exploration, space resource extraction, and national defense applications. The structured approach adopted here is derived from the methodology outlined by Notar and Cole in their 2010 paper, 'Literature Review Organizer,' (Notar & Cole, 2010) ensuring a comprehensive exploration of the subject, from foundational theories to practical applications, and the challenges and opportunities they present.

**Theoretical Background**

The concept of Lagrange points is a well-studied subject in celestial mechanics, offering insight into the complex gravitational interactions between two large bodies and a much smaller one of negligible mass. Named after the Italian French mathematician Joseph-Louis Lagrange, who made significant contributions to their understanding in the 18th century (Fraser, 1983), these points mark positions in space where the gravitational forces of two large bodies, such as the Earth and the Moon, balance the centripetal force felt by a small object, such as a satellite. The equilibrium allows the smaller object to maintain a stable relative position.

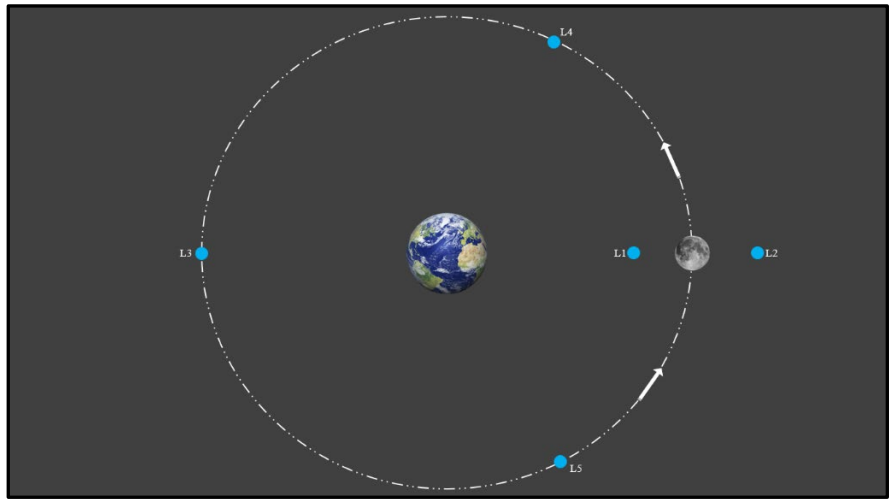

Figure 1 – Diagram illustrating the concept of Lagrange points using the Earth-Moon Lagrangian points. Created by the author.

There are five such points, labeled L1 through L5, each with unique characteristics and implications for space exploration. The L1, L2, and L3 points lie along an imaginary line connecting the centers of the two large bodies, where L1 is positioned between the two main bodies (but not necessarily in the middle), L2 lies on the outside of the smaller of the two larger bodies, and L3 is along the orbital path of the smaller of the two larger bodies at 180 degrees ahead or behind the smaller body. For this paper, L1, L2, and L3 will be referred to as the "collinear points" and are the lesser stable points where objects at those points have their orbital velocities degrade over time.

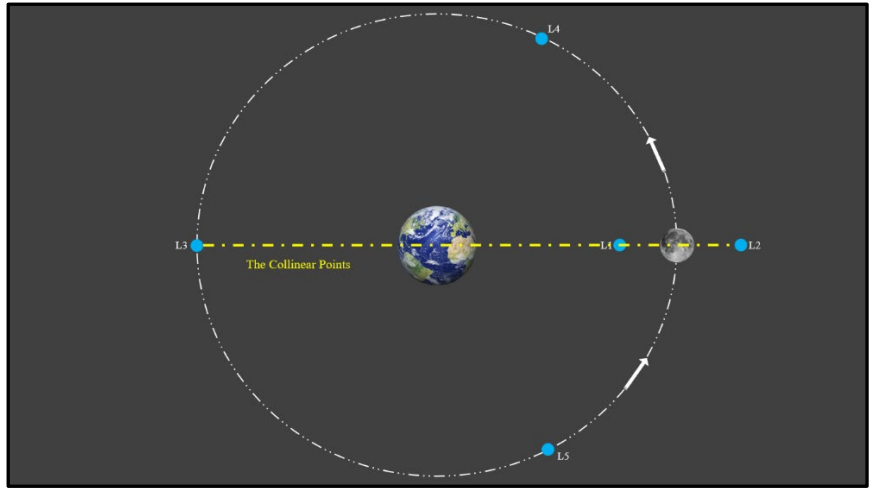

Figure 2 – Diagram illustrating the Collinear points using the Earth-Moon Lagrangian points. Created by the author.

On the other hand, L4 and L5 form the apexes of two equilateral triangles, with the two large bodies at their vertices, and are significantly more stable than their collinear counterparts. L4 is 60 degrees ahead of the smaller body along its orbital path and L5 is 60 degrees behind the smaller body, trailing its orbital path. For this paper, L4 and L5 will be referred to as the "triangular points" and noted for their increased stability compared to the collinear points.

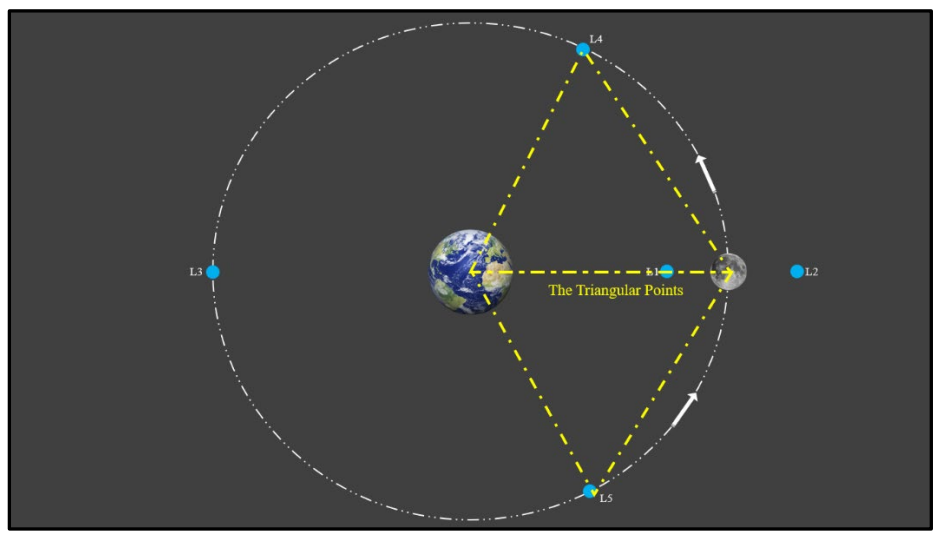

Figure 3 – Diagram illustrating the Triangular points using the Earth-Moon Lagrangian points. Created by the author.

For a simplified mathematical overview of the locations of all five Lagrange points, we will discuss the relevant theories through the lens of the five Lagrange Points in the Earth-Moon System (Bairwa, Pal, Kumari, Alhowaity, & Abouelmagd, 2022) and the mathematical frameworks that describe these unique spatial areas (Parker & Anderson, 2014). Calculating the exact location of Lagrange points involves solving the equations derived from the gravitational force equilibrium conditions in the circular restricted three-body problem (CRTBP) (Curtis, 2013). Here, we will outline a simplified version of the mathematics involved in locating the L1, L2, and L3 points for the Earth-Moon system, focusing on the concept rather than a rigorous derivation. For the collinear points, the forces involved include the gravitational forces exerted by the two main bodies and the centrifugal force due to the rotation of the system.

*Collinear Assumptions and Simplifications*

1. The two main bodies (the Earth and the Moon) are in circular orbits around their common center of mass.
2. The mass of the third body (a satellite) is negligible compared to the two main bodies.
3. We are considering a straight-line model for L1, L2, and L3, which lie along the imaginary line connecting the centers of the two larger bodies.

*Collinear Notations*

- $M_1$: Mass of the larger body (Earth)
- $M_2$: Mass of the smaller body (Moon)
- $R$: Distance between the centers of the two main bodies.
- $r$: Distance from the smaller body to the Lagrange point.
- $G$: Gravitational constant.

*Calculating L1*

The L1 point is located between the two main bodies (but not necessarily in the middle of that distance.) The gravitational forces from both bodies and the centrifugal force (which depends on the angular velocity of the system) all act on the satellites. The equation to find $r_{L1}$, the distance of L1 from the smaller body (Moon), can be simplified into a form of the equation that is solvable but requires numerical methods for precise solutions. The simplified equation for the location of L1 can be approximated by:

$$\frac{M_1}{(R-r)^2} = \frac{M_2}{r^2} + \frac{M_1}{R^2} - \frac{r(M_1 + M_2)}{R^3}$$

Source: *Lagrangian Point,* (2024)

*Calculating L2*

The L2 point lies outside the orbit of the smaller body (on the opposite side of L1.) The formula for $r_{L2}$, the distance from the smaller body to L2, is very similar to L1 but the balancing of the centrifugal effect is arrived by the gravitational force of the two larger bodies. The simplified equation for the location of L2 can be approximated by:

$$\frac{M_1}{(R+r)^2} + \frac{M_2}{r^2} = \frac{M_1}{R^2} + \frac{r(M_1 + M_2)}{R^3}$$

Source: *Lagrangian Point,* (2024)

*Calculating L3*

The L3 point is located on the line defined by the two main bodies but beyond the larger body (opposite the smaller main body.) The equation for L3 involves a similar balance of forces but is situated such that it almost mirrors the position of L2 with respect to the larger body.

$$\frac{M_1}{(R-r)^2} + \frac{M_2}{(2R-r)^2} = (\frac{M_1}{M_1 + M_2}R + R - r)\frac{M_1 + M_2}{R^2}$$

Source: *Lagrangian Point,* (2024)

*Triangular Assumptions and Simplifications*

1. The two main bodies (The Earth and the Moon) are in circular orbits around their common center of mass.
2. The mass of the third body (satellites) is negligible compared to the two main bodies.
3. The L4 and L5 points form equilateral triangles with the two main bodies.

*Triangular Notations*

- *a* represents radial acceleration.
- *r* refers to the distance from the large body M1
- *sgn* (x) is x's sign function.

*Calculating L4 and L5*

The locations of L4 and L5 are determined geometrically rather than through the kind of balance equations used for the collinear points. They form equilateral triangles with the two main bodies and do not have a simple 'distance from' formula. The mathematical representation by making use of radial acceleration is below:

$$a = \frac{-GM1}{r2} sgn(r) + \frac{GM_2}{(R-r)^2} sgn(R-r) + \frac{G\big((M_1 + M2)r - M_2 R\big)}{R^3}$$

<div align="right">Source: *Lagrangian Point,* (2024)</div>

**Areas of Focus**

Following the theoretical groundwork laid in the preceding section, our exploration transitions to the practical realms where the concepts of Lagrange points are applied. These applications span diverse yet interconnected fields, each representing a unique "Area of Focus" that collectively illustrates the multifaceted potential of Lagrange points in advancing human knowledge and capability in space. The subsequent sections are dedicated to four primary areas: astronomical research, space explorations, space resource extraction, and national defense. Each domain showcases the strategic importance and utility of Lagrange points beyond theoretical constructs, into tangible, impactful applications.

In addition to the specified "Areas of Focus," it is important to acknowledge a fundamental application that intersects with each of these domains: the use of Lagrange points for communication satellites or relays. The strategic positioning of communication infrastructure at these points can significantly enhance data transmission capabilities across vast distances, offering a continuous and reliable link between Earth and the farthest reaches of our Solar System. Whether supporting astronomical research by transmitting unprecedented volumes of scientific data back to Earth, facilitating real-time communication for deep-space exploration missions, providing the backbone for logistics in space resource extraction, or enhancing the network resilience in national defense applications, communication satellites placed at Lagrange points serve can weave together many space endeavors.

Relay satellites placed at various Lagrange points in the inner Solar System are discussed at length by Howard & Seibert (2010) as a proposal for eliminating communication outages during Solar Super Conjunction, but in reality, the proposed models could be applied to almost any endeavor described in our "Areas of Focus." This omnipresent application underscores the interconnectedness of the research areas, reinforcing the notions that advancements in communication technologies at Lagrange points are indispensable for the success and sustainability of future space activities.

### *Astronomical Research Applications*

Astronomical research represents one of the most extensively explored and currently operational applications of Lagrange points. These unique orbital positions offer unparalleled opportunities for space observatories to conduct uninterrupted studies of the cosmos, free from Earth atmospheric distortions. While it is challenging to comprehensively catalog every piece of literature that discusses or utilizes Lagrange points for astronomical purposes, the selected studies highlighted in this section aim to encapsulate the major themes and breakthroughs in this domain. These examples underscore the critical role Lagrange points play in advancing our understanding of the universe, showcasing their value not just in theoretical research but in practical, ongoing scientific inquiry.

In 1978, the National Aeronautics and Space Administration (NASA) in partnership with the European Space Agency (ESA) launched the third of three International Sun-Earth Explorer (ISEE) satellites. This third satellite would be the first human-made object to visit and reside in a Lagrange point (Farquhar, 2001). Placed at the Sun-Earth L1 point, some 1.5 million kilometers from Earth and toward the Sun, ISEE-3 (and later ISEE-3/ICE for International Cometary Explorer) was a monumental craft not only for its visit to the Sun-Earth L1 point, but to also be the first satellite to make a study of Earth's magnetosphere and the first to fly past a comet. It is entirely appropriate then that this literature review first and primarily study the astronomical research applications of the Lagrange points in the Solar System.

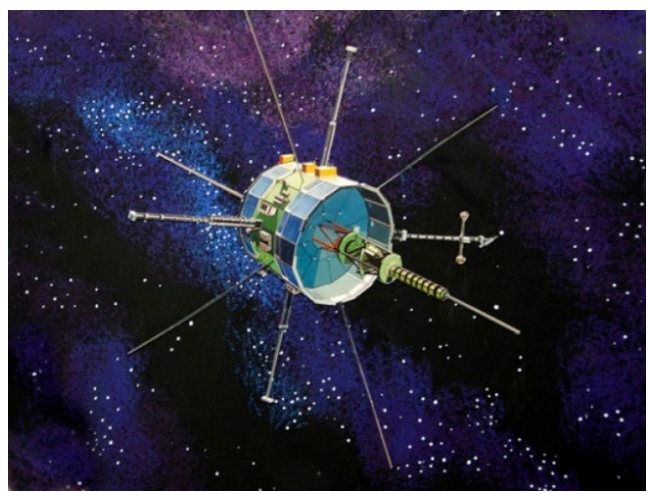

Figure 4 – An artist's impression of NASA's ISEE/ICE spacecraft, the first manmade object to visit a Lagrange point, the Sun-Earth L1. Source: *NASA,* (2024)

The Sun-Earth Lagrangian system offers several of the most discussed and exploited Lagrange points. Besides never having the Sun obstructed by the Earth or Moon equally, observations of the Sun and solar weather can be made before those solar weather impacts are felt on Earth. The Sun-Earth L5 point is a proposed location for space weather research by Vourlidas in their 2015 research into heliophysics, or the physics of the sun (Vourlidas, 2015). A satellite trailing Earth's orbit is the equivalent of being in geosynchronous orbit around the Sun. Here, imaging of solar activity, specifically solar wind from coronal mass ejections (CME) can be performed at least three days prior to any impacts or effects on Earth. CMEs were the subject of the Solar Terrestrial Relations Observatory (STEREO) that the National Aeronautics and Space Administration placed at the Sun-Earth L4 and L5 positions (Driesman, Hynes, & Cancro, 2008). Here, the identical satellites offered a stereoscopic image of the Sun and its condition. This same concept using the Sun-Earth L4 and L5 points to make heliospheric studies was further explored by Bemmporad, expanding on the STEREO program with the additional area of study on coronal magnetic fields (Bemmporad, 2021). Complimenting Vourlidas' research with a theoretical spacecraft at the Sun-Earth L1 point between the Sun and the Earth, Hapgood discusses a system of solar focused satellites, one at L1 and another at L5, that could further advance the study into space-weather generated by the Sun and the impacts on Earth (Hapgood, 2017). The exploitation of the Sun-Earth L1 position has been in use since as early as 1995 with the European Space Agency (ESA) and NASA's Solar and Heliospheric Observatory (SOHO) satellite that has collected an abundance of data on the Sun including solar wind and the solar corona (Roberts, 2003). Beyond its original mission of solar observations, SOHO has also discovered at least 4,000 comets as of the writing of this review (Frazier, 2020). The Sun-Earth L2 point hosts one of the most prominent and discussed modern astronomical observatories, the James Webb Space Telescope (JWST) (Menzel, et al., 2023.) Here, unobstructed by the Earth's atmosphere and other gravitational effects, JWST produces some of the clearest and highest definition images of the outer Solar System and countless celestial bodies and objects far beyond the Solar System.

Reaching further into the Solar System at the location of the Sun-Jupiter Lagrangian system, NASA's Lucy spacecraft is currently in orbit of the gas giant, with a mission to study various of the multitude of natural objects found at the Sun-Jupiter L4 and L5 positions (Levison et al., 2021). Located here are an innumerable number of asteroids that may be remnant material from the earliest age of the Solar System, which Lucy will study to provide possible insight into the conditions present at the birth or infancy of the Solar System.

*Astronomical Research Application Challenges and Opportunities*

In the astronomical research area of focus, the Sun-Earth Lagrange points were discussed, and the literature reviewed primarily covered solar research or studies focusing on solar weather and how it may impact the Earth. Indeed, the Sun-Earth L1, L2, L4, and L5 positions are measurably easier to deliver spacecraft or satellites to, and the Sun is easily observed from most of these locations. Not only are there more proposals for applications studying the Sun from these positions, but the extent of the currently published literature would also consume an entire thesis and literature review itself. Presently, the Sun-Earth L2 positions hosts or is proposed to host observatories that are designed to observe the outer Solar System or the space beyond the Solar System. Nothing could be retrieved or reviewed on any proposal of the Sun-Earth L3 point, the position 180 degrees opposed to Earth's present position along its orbital path. It remains to be discussed if there is any use for this position that could not be adequately resolved using the L1, L4, or L5 positions. The Sun itself would likely be the principal subject of any such observatory here, and while it may offer insight into the solar weather and fluctuations on the unobserved side of the Sun, any result or data gathered may not be insightful enough to warrant the associated cost with sending an astronomical observatory to such a distance.

      The literature highlighted the use of the deepest astronomically focused observatory at any Lagrange point in the Solar System, NASA's Lucy at the Sun-Jupiter L4 and L5 positions. Launched in 2021, Lucy only recently reached the Jovian system, and due to the enormous distance from Earth and the distances between Jovian L4 and L5 points, it is expected to take at least twelve years for Lucy to complete its mission and return any data back to Earth. The challenge in dealing with Lagrange points this far out in the Solar System comes from the time associated with delivering spacecraft to this depth and the risk that may be associated with technical malfunctions that would be well beyond the reach of humans to physically correct. Yet, the Jovian L4 and L5 points offer some of the most tantalizing opportunities with its Trojan and Greek asteroid populations. As the literature discussed, these asteroids may have been present at the earliest age of the Solar System and never accumulated into a moon or body themselves or amassed into a larger planet or moon in the Solar System. The potential to study this material may offer insight into the conditions present at the earliest point in the Solar System and the materials ejected or collected during the chaotic early formation.

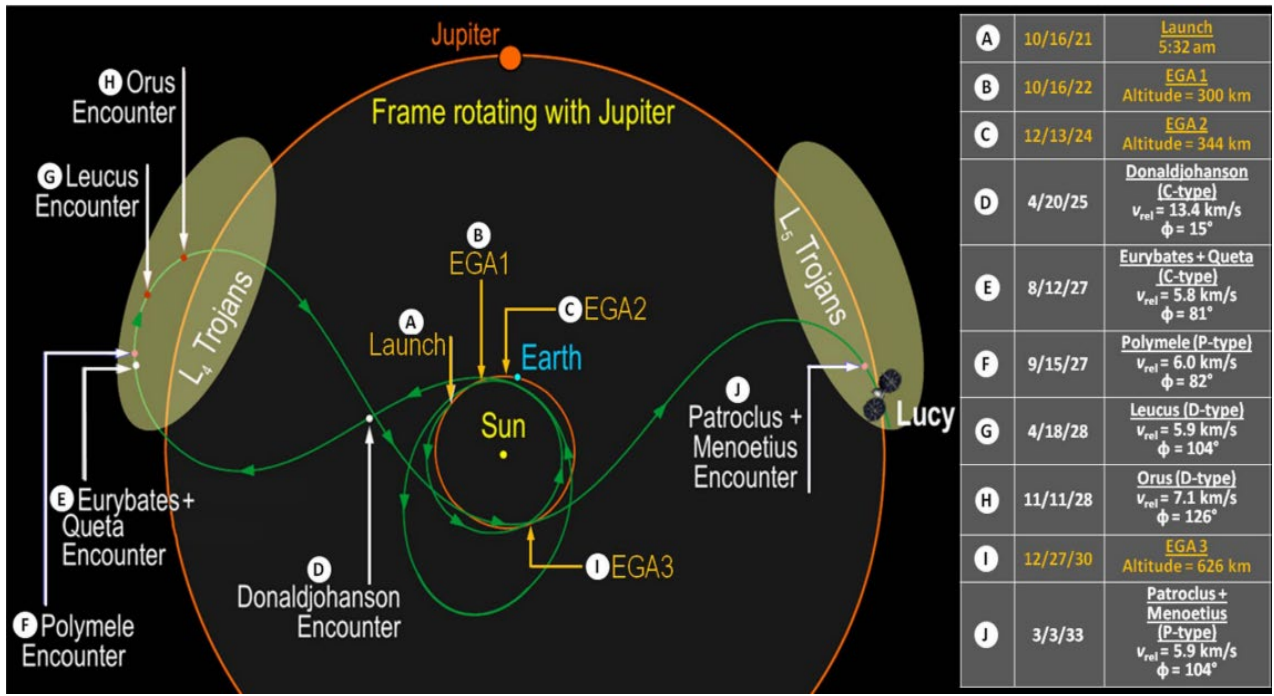

Figure 5 – Trajectory of Lucy to the Sun-Jupiter L4 and L5 points.

Source: *Levison, Olkin, Noll et al.* (2021)

There exists a significant gap in the literature for astronomical research using Lagrange points outside of the Sun-Earth and Sun-Jupiter systems. Nothing could be found for any proposal for the Sun-Mercury, Sun-Venus, Sun-Mars, Sun-Saturn, Sun-Uranus, or Sun-Neptune systems. Some research did uncover the potential for studying trans-Neptunian objects, or those objects beyond the orbit of Neptune (Barucci, Boehnhardt, Cruikshank, & Morbidelli, 2008), but no proposal for what type of spacecraft or observatory might be best suited for this study or at which, if any, Sun-Neptune Lagrange point would be most optimal. There does exist a family of "Neptune Trojans" or asteroids within the Sun-Neptune L4 and L5 positions (Lin, et al., 2016) that Horner and Lykawka postulate could be a source for the Centaurs, the small comet-like bodies that orbit the Sun in the space between Jupiter and Neptune (Horner & Lykawka, 2010). The potential here to study other primordial bodies that existed in the early stages of the Solar System is like that of the Jovian Trojans but could also lend further insight into the orbital mechanics and dynamics of the outer Solar System. No proposals could be found discussing the several other Lagrange points that exist between the gas giants (Jupiter and Saturn) and ice giant planets (Uranus and Neptune) and their companion moons. Just as the multitude of satellites at the Sun-Earth systems that are used to study the Sun, observatories positioned at any of the L1 positions existing between the outer Solar System planets and their respective moons could offer an unobstructed, long-term, and in-depth study of that major planet.

*Space Exploration*

The use of Lagrange points facilitates space exploration by providing strategic positions for spacecraft to reduce fuel consumption and optimize mission trajectories. One study focused on the Sun-Earth and Earth-Moon Lagrange points, analyzing trajectories to these points and their potential for accessing the surface of celestial bodies (Baoyin & McInnes, 2006). Other research in the use of the L4 and L5 points by Antonio Prado explains in detail how the collinear L1, L2, and L3 points are more unstable than their triangular L4 and L5 counter parts, making them more a more ideal location for a space-station, especially in the Sun-Earth Lagrange system (Prado, 2006). Another study explores the use of the Earth-Moon L1 and L2 positions as potential locations for rendezvous and docking operations for crewed or robotic missions supporting further Lunar exploration (Lizy-Destrez, Beauregard, Blazquez, Campolo, Manglativi, & Quet, 2019). These sites could potentially support the NASA-led Lunar Orbital Platform-Gateway (Duggan, Simon, & Moseman, 2019). The Earth-Moon L1 point is further explored as a dedicated service point for a gateway type stepping-stone for Lunar surface missions (Yazdi & Messerschmid, 2008). Still, others suggest the use of the L2 point as a location for astronauts to operate telerobotic missions on the Lunar surface that could land, collect, and return samples from the far side of the Moon that is tidally locked, facing away from the Earth (Burns, Kring, Hopkins, Norris, Lazio, & Kasper, 2013).

Several works have studied the use of both the collinear and triangular Lagrange points of the Earth-Moon and Sun-Earth systems as launching points for deeper space missions due to the gravitational equilibrium they possess (Morimoto, Yamakawa, & Uesugi, 2007). Several of these papers reference what appears to be one of the earliest research papers into the exploitation of the Earth-Moon L1 and L2 points for both Lunar communication as well as spacecraft staging points (Farquhar, 2012). The concept of establishing "Lagrangian Bases", or long-term crewed stations at the Sun-Earth Lagrange points, are explored in depth by another work that proposes not only different mission scopes that could be supported by such bases, but also different base designs and configurations (Sehgal, 2023). Here, the gravitational stability of the Lagrange points is the most referenced reason for their proposed locations and the bases could serve as rendezvous, refueling, or operational bases supporting crewed deep-space missions to the outer Solar System.

Only one study could be found to propose the use of the Sun-Mars L1 position as a point to facilitate the exploration of the Red Planet (Eapen & Sharma, 2014). Here, a combination of parking orbits in near-Earth Cislunar orbit, the Sun-Earth L2 point, and finally the Sun-Mars L1 point could reduce the changes in velocity (and thus reduced fuel consumption) and increase the payload capacity for Martian interplanetary exploration. Another study explored the use of communication satellites at the Sun-Mars L1 and L2 points to better facilitate Martain exploration (Strizzi, Kutrieb, Damphousse, & Carrico, 2001). The concepts discussed by Strizzi, and colleagues are not unique to Martian exploration and could be applied to almost any planetary surface exploration, which supports this literature reviews intent of not covering communication applications separately, but as a continuous application that supports almost every Area of Focus discussed.

*Space Exploration Challenges and Opportunities*

The literature presents several, in fact too many to be referenced in totality, options for the exploitation of the Sun-Earth and Earth-Moon Lagrange points for the exploration of the Lunar surface and the Cislunar space between the Earth and Moon. In some arguments, the collinear L1 and L2 positions are described as unstable and therefore undesirable, preferencing instead the L4 and L5 points. Then others argue that the L1 and L2 positions may be more ideal for conducting short-term rendezvous to and from the Lunar surface. It is without question that the concept of the Earth-Moon Lagrange points has been comprehensively examined as launching or staging points for further Lunar exploration. Still, no study outright mentioned the use of the Earth-Moon L3 point except to include in with its collinear counter parts L1 and L2.

Surprisingly, only one research team could be found to have explored the use of the Sun-Mars Lagrange points for Martian exploration. The discussion of sending crewed mission to Mars to accompany the smattering of robotic mission already to land on and explore the Martian surface has been increasing in recent years, with one study outright exploring the concept of human outposts on Mars (Gruenwald, 2014). It would appear then that an opportunity for the exploitation of the Sun-Mars L1, L2, L4, and L5 points would be ripe for further understanding. Furthermore, just as this study proposed the Sun-Earth L2 point as a parking orbit to better align with the Sun-Mars L1 point, it may be equally advantageous to utilize the Sun-Mars L2 point for deeper space missions to the asteroid belt and gas giants of the outer Solar System.

*Space Resource Utilization Applications*

This area explores the potential for using Lagrange points as staging areas for asteroid mining and other space resource utilization efforts. It considers the logistical, technological, and economic aspects of such ventures. Here again, the state of the current literature is primarily focused on the inner Solar System, or those Lagrange points that can be found at the orbit of Mars and inward toward the Sun.

Perhaps the most abundant resource in the Solar System is the energy emitted by the Sun. The Sun itself accounts for the predominance of mass in the Solar System, and its energy output is greater than anything in the Solar System. The first study reviewed utilizing that solar resource to maintain station-keeping for satellites or spacecraft in the collinear L1 and L2 points in the Earth-Sun system (Bookless & McInnes, 2008). As previously discussed, the collinear points are the more unstable points of the five, with the triangular L4 and L5 points of any system being more stable. The authors argue that the use of solar sail propulsion to stabilize the orbits of satellites or craft in the L1 or L2 points would reduce the fuel requirements for station-keeping at those points and allow for more payload capacity.

Instead of utilizing the Sun as source for more energy, the next study reviewed the use of the Sun-Earth L1 point as place to position a cloud of satellites that could block 1.8% of the solar flux to cool the Earth to combat dangerous climate changes (Angel, 2006). Satellites at this point would be positioned at a stable and yet critically important point to have this effect. It might even be argued then that the resource exploited here is not solar energy, but rather the gravitational stability inherent to the Lagrange point itself, which is further supported by previously cited research that exercised several Lagrange points for their ability to station spacecraft for long durations for both astronomical and exploration missions without the use of a large volume of fuel required to maintain their position.

Beyond the gravitational and solar energy potential, the Solar System is almost littered with an abundance of asteroids, comets, and other smaller objects potentially rich in minerals and metals that are not readily available on Earth. Furthermore, the existence of materials used on Earth that can be found in these smaller bodies that are used to construct spacecraft components or even fuel offers the potential for the extraction and use of the material to extend the lifespan or mission scope of spacecraft at great distances from Earth. The first study in this review that explored that potential argued the use of the Sun-Earth L2 point, a safe distance away from the Earth and the Moon, as a capture point for Near Earth Objects (NEOs) where said resources could be mined, extracted, or studied and where other locations, such as within the Earth-Moon system, could be reached at relative ease (Llado, Ren, Masdemont, & Gomez, 2012).

NASA itself has made a proposal for a craft to visit the Sun-Earth L4 and L5 positions to explore the objects that have been identified in these stable points (John, Graham, & Abell, 2015). The proposal, published through the Johnson Space Center argues for the use of these accessible target locations where research into the microgravity environments around objects just a few hundred meters in diameter could provide useful information on future crewed and robotic exploration missions. Adjacent to this research is the possible mining of water and other resources present in Neath Earth Asteroids (NEAs) within Cislunar space that could be utilized by astronauts transitioning to the Sun-Earth L2 point (Calla, Fries, & Welch, 2019), where they would be at a significant distance from any *in situ* research. In this proposal, it is the object or crew at the L2 point that is benefiting from the resources nearby. Another study supports this proposed method of using NEOs and NEAs as a source of water for the long-term sustainment of astronauts at the collinear points in the Sun-Earth system (Vergaaij, McInnes, & Ceriotti, 2021). At these locations, the space resources in Sun-Earth, Sun-Moon, and Cislunar space become the most attractive and most feasible sources of essential resources outside of Earth itself. The article does go further to argue that crewed missions on the surface of Earth, the Moon, and Mars could likely extract water and other resources more efficiently from those bodies themselves.

As NASA and private space agencies continue to look forward to crewed missions on the surface of Mars, Space Resource Utilization (SRU) or *in situ* resource use becomes more relevant. One study in 2018 proposed a method of capturing and moving asteroids into the Sun-Mars L1 and L2 positions to be mined or broken down into usable material supporting surface missions on the Red Planet (Tan, McInnes, & Ceriotti, 2018). Here, the instability of the collinear points is less of a concern as it would appear. The asteroids would only remain in those positions for as long as it took to consume them, a far shorter period than it would take for their orbital stability to degrade due to the inherent instability of a collinear Lagrange point, as discussed previously.

*Space Resource Utilization Applications Challenges and Opportunities*

The literature for this Area of Focus primarily discusses the asteroid and small Solar System body resource exploitation to facilitate crewed or robotic space exploration missions near Earth or as far out as Mars. Multiple works highlight the feasibility of and ease of access to the objects in the Sun-Earth L4 and L5 points and propose actual research or utilization of the bodies therein. The opportunity here is the relatively proximity of those points to Earth and the level of spacecraft technology required to reach them that some agencies may have access to now or soon. There appear to be no shortages of NEAs or NEOs that are available to be either studied on site at L4 and L5 or can be reasonably guided to those points or to the Sun-Earth collinear points for further study or use. The opportunity to maneuver some asteroids or other smaller bodies into the Sun-Mars Lagrangian system is also being reviewed, with the mining or exploitation of which would support the exploration of Mars. Interestingly, the largest source of raw material in the inner Solar System – the main asteroid belt between the orbits of Mars and Jupiter – is only briefly discussed in one paper as to acknowledge its applicability for exploitation after asteroids and other objects between Earth and Mars orbits are exhausted. This is likely due to the enormous distance between Mars and the asteroid belt, at which point the utilization of asteroid belt resources may be more pertinent to a continued mission onward to Jupiter and the outer Solar System than back in. Research into crewed missions that might require those resources to venture outward from Martian orbit is limited as the technology to support such quests is currently beyond the scope of NASA and other space agencies.

*National Defense Applications*

This section addresses the strategic importance of Lagrange points in national defense, including surveillance, space situational awareness, and other security-related uses. Emphasis is on the Earth-Moon Lagrange points and Cislunar space and relevant open source, unclassified, information from US government and defense agencies. Naturally, as it relates to those military or defense agencies on Earth, the exploitation of local Lagrange points, as in the Sun-Earth and Earth-Moon points, are covered.

Space Domain Awareness (SDA) is defined by the United States Space Force (USSF) as the ability to rapidly detect, warn, characterize, attribute, and predict threats using space-based and other assets (Space Systems Command, 2024). The first work reviewed for this section discusses not the Earth-Moon Lagrange points themselves, but proposed orbits used to monitor those points along with the rest of Cislunar space (Wilmer, 2021). Here, periodic orbits within Cislunar space to include orbits out to and around all three Sun-Earth collinear points is discussed as means to improve SDA through monitoring of all significant points within Cislunar space to include the Earth, the Moon, and the Lagrange points.

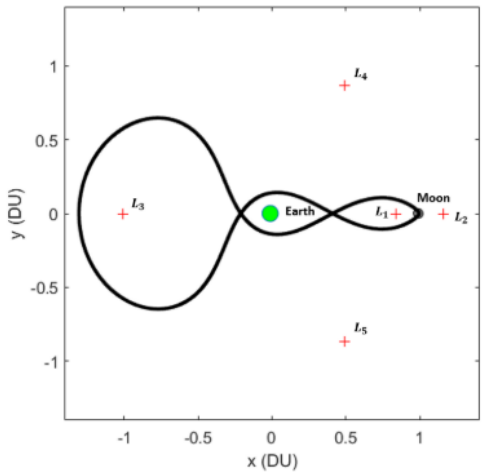

Figure 6 – Example Cislunar Periodic Orbits around the Earth Moon Lagrange System.

Source: *Wilmer* (2021)

The Chinese Queqiao communication satellite currently resides at the Earth-Moon L2 point (Zhang, 2021), and the author makes note that the continued exploitation of the Earth-Moon Lagrange points by national defense agencies is likely to increase in the coming years. Zhang discusses the intended use of the satellite for relaying communications from Earth to a Chinese lunar rover on the surface of the Moon and how its success is shaping the development of future lunar relay systems at the Earth-Moon Lagrangian system.

Military communication satellites are discussed further in *Eyes on the Prize: The Strategic Implication of Cislunar Space and the Moon* (Kaplan, 2020). Here the author details the use of the Earth-Moon L1 and L2 points, but also theorizes how both triangular and collinear points in the Earth-Moon system could be exploited for the "long-term parking" of military satellites to observe not on the Moon and Cislunar space, but the Earth as well. The author further addresses the use of possible space elevators at the Earth-Moon L1 and L2 points to reduce the cost of deploying personnel and equipment from lunar orbit to the surface and the extraction of material from the surface back into orbit.

While more in the realm of planetary defense than national defense, Maccone (2004) proposes the use of the Earth-Moon L1 and L3 points as holding points for space-based missiles intended for deflecting or destroying Near Earth Objects (NEOs) like asteroids that may pose a threat to Earth. The author argues that missiles fired from these locations, rather than from the Earth's surface, would provide more favorable trajectories for any remaining debris from a destroyed asteroid that may still impact the Earth after the bulk of the material is destroyed with the original impact.

*National Defense Applications Challenges and Opportunities*

The discussion of the use of Lagrange points for military actions or defenses is a sensitive topic. Despite its sensitivity, the leading space powers, primarily The United States and China, are already establishing Cislunar and Lagrangian architectures of space-based systems that may presently be used for peaceful and scientific purposes but could readily be converted into military applications. There appears to be no shortage of opportunities for exploitation as all five points are reviewed with potential applications in the Earth-Moon Lagrangian System, including the L3 point, which almost every other potential Area of Focus seems to not find an applicable use for. Here the challenge will be in Space Policy and Space Diplomacy, something this literature review does not have the scope to cover, but is a separate subject that others are writing more deliberately about (Hall, 2007; Committee, 2016; Department of Defense, 2011).

# CHAPTER THREE: THEORETICAL FRAMEWORK

**Chapter Introduction**

Recent advancements in space technology and exploration have spotlighted the strategic importance of Lagrange points in the solar system (Bookless & McInnes, 2008; Duggan et al, 2019; Farquhar, 2012). Despite the increasing number of missions exploiting these points (Gruenwald, 2014; Hapgood 2017; Howard & Seibert, 2010), a comprehensive understanding of their full potential remains fragmented across various space research domains. This study seeks to consolidate knowledge surrounding Lagrange point applications and identify gaps in current literature, particularly in the systematic methodological review of their utility across different space missions. Grounding our investigation in Systems Theory provides a holistic framework to examine the interconnectedness of space missions utilizing Lagrange points, offering insights into optimizing their use for future endeavors.

**Justification for Systems Theory**

Systems Theory, with its roots in understanding complex entities as holistic systems rather than merely the sum of their parts (Skyttner, 2002), offers a robust framework for analyzing the multifaceted applications of Lagrange points in space exploration. This theory allows for the examination of how individual components of space missions (e.g., spacecraft deployment, scientific research, communication networks) integrate and interact at Lagrange points to produce emergent behaviors and outcomes. Its application in space mission analysis has precedent (Howell, Barden, & Lo, 2020), providing a validated approach for studying non-linear, complex systems, which is characteristic of space exploration missions. The Systems Theory's emphasis on interrelationships and interactions (Bailey, 2001) makes it apt for exploring how different missions leverage Lagrange points, thereby filling the identified literature gaps.

**Summary of Systems Theory**

Systems Theory posits that to fully understand a system's functionality and potential, one must consider the system in its entirety, including all interaction between its components (Skyttner, 2002). Within the context of space exploration, each mission utilizing a Lagrange point can be seen as a system component with specific objectives (e.g., Earth observation, deep space exploration). By applying Systems Theory, this research will map out these components' interactions, highlighting synergies and potential areas for optimization. A diagrammatic representation of the theory applied to Lagrange point missions will be presented, highlighting the interconnections and flow of information and resources between different missions.

**Development of Hypotheses**

Based on the theoretical framework of Systems Theory, the following hypotheses are proposed for testing in this study:

1. **H1:** Missions utilizing Lagrange points exhibit a high degree of systemic interdependence, influencing their design and operational strategies.
2. **H2:** The integrated analysis of missions at Lagrange points, grounded in Systems Theory, reveals optimization pathways that are not apparent when missions are analyzed in isolation.

These hypotheses will guide the empirical investigation, focusing on how the systemic interactions between various missions can enhance the overall utility of Lagrange points in space exploration.

**Establishing Construct Validity**

The application of Systems Theory to the study of Lagrange point applications in space missions establishes construct validity (Broniatowski & Rucker, 2018) by ensuring that our research methodology aligns with the complexity and interconnected nature of the subject. By framing our study within this theory, we can validly interpret how different components of space missions at Lagrange points interact, complement, and impact each other, providing a comprehensive understanding that matches the real-world complexity of space exploration.

**Systems Theory in Action: Lagrange Point Utilization for Human Mars Exploration.**

At the European Geosciences Union Assembly in 2023, Deputy Direct of the NASA Office of JPL Management and Oversight, Dr. Azita Valinia, then the Chief Scientist at NASA Engineering and Safety Center, presented a proposal for advanced reconnaissance missions needed for the human exploration of Mars (Valinia, 2023). This presentation referenced the report made by Dr. Valinia and her collaborators *Safe Human Expeditions Beyond Low Earth Orbit* (Valinia, Allen, Francisco, Minow, Pellish, & Vera, 2022).

In this presentation, Dr. Valinia highlighted the need for high resolution imaging of the Martian surface for landing site reconnaissance, base construction, and in-situ resource utilization. Further planetary observation could assist in surface weather prediction that will influence surface operations and ascent and descent operations. Martian surface weather is not the only planning factor, however. Space weather observations and predictions will be critical for ensuring the health and safety of future Martian astronauts en route to, in orbit of, or on return missions from the red planet. A robust communication and navigation system of satellites will be essential to the planning and execution of a human exploration of Mars and would need to be in place prior to the first astronaut ever leaving the surface of Earth.

Dr. Valinia cited the use of the Sun-Mars Lagrange points for some of these applications, particularly the space weather observations, but through an application of some of the other research previously cited in this study, those same Lagrange points could be part of the interconnected system supporting future missions, specifically the communication system and the transfer of astronauts to and from the surface of Mars.

**Application of Systems Theory**

*Interconnectivity*

A principal component of Systems Theory is the interconnectedness of the components of the system and the behavior and relationship of those components to the entirety of the system (Meadows, 2008). The infrastructure of the mission supporting the human exploration of Mars will inherently be a dynamic collection of systems serving multiple functions creating the framework by which humans will move to, from, and explore Mars and the method by which they communicate back to Earth. Key pieces of terrain that mission planners will need a thorough understanding of and efficient exploitation of to make the human exploration of Mars successful include Lagrange points, along with Earth orbits, Cis-lunar space, Lunar orbits, Martian orbits, and the trajectories between those various orbital altitudes – all of which have been discussed previously in this study.

As Dr. Valinia proposes, space weather satellites at the Sun-Mars L1, L4, and L5 positions provide the predictability required for efficient and safe mission planning during optimal solar cycles to decrease the exposure of solar radiation as much as can be tolerated for a mission with such high inherent risk. This theory is sound and is supported by other such space weather and solar weather studies already proposed and operating in Sun-Earth Lagrange points like NASA's STEREO and ISEE-3/ICE satellites (Driesman et al., 2008; Farquhar, 2001) and other proposals that support similar functions previously cited in this study (Vourlidas, 2015; Hapgood, 2017; Roberts, 2003; and Bemporad, 2021).

Further, the previously cited Lagrange communication relay system (Strizzi et al., 2001) could also take advantage of the Sun-Mars Lagrange system that could support Dr. Valinia's need for a robust communication network of satellites that will serve as the primary method of communication between Martain astronauts and Earth-based observers and administrators. That study is supported with a report (Howard & Seibert, 2010) that proposes a relay of Lagrange-based communication satellites at various Lagrange points in the inner solar system that could provide a redundant capability in the event of Solar Superior Conjunctions. This proposed network would not only support Martain exploration, but all human exploration happening off the Earth's surface.

Finally, the ascent and descent of astronauts to the Martain surface could be facilitated using the Sun-Mars L1 points as previously proposed in the Lunar Gateway review that utilizes the Earth-Moon L1 point to transfer not only personnel but resources to and from the surface of the Moon (Burns et al., 2013).

An application of at least fifteen Lagrange points between Earth, the Moon, and Mars is one of the many interconnected nodes and systems that support the first astronauts to explore Mars in detail.

*Feedback Loops*

Systems Theory discusses feedback loops, or how the predicted and unpredicted behavioral changes and influences of components within a system influence the design and improvement of a system over time (Meadows, 2008). Here, the existing space exploration or astronomical observations systems that utilize Lagrange points could provide critical points of feedback that will influence the eventual architecture supporting Martian exploration. The forthcoming Lunar Gateway will provide essential feedback on the frequent delivery and recovery of personnel, equipment, and resources to and from the surface of a celestial body with a thin atmosphere and lower-than-Earth gravitational well, a practice that stakeholders in this eventual mission will not have much experience with prior to the execution of the first several missions.

*Adaptability*

A core tenant of Systems Theory emphasizes the capacity of a system to adjust its functioning in response to changes in its environment or in parts of the system itself (Meadows, 2008). In the context of supporting the human exploration of Mars, adaptability manifests in several key dimensions including technological evolution, mission objectives, and environmental unpredictability.

The locations of the various Lagrange points between Earth and Mars can facilitate a rapid change in the advancement of propulsion systems that will be utilized by the spacecraft carrying personnel and equipment to and from Mars. As previously cited (Parker, 2014), their gravitational stability makes them ideal parking space for the optimal conjunction of the planets or moons to facilitate the transfer of such spacecraft. These points will move closer and further away from each other during regular orbital processions creating opportunities for less-advanced and more-advanced propulsion systems to exploit different travel requirements.

As discoveries are inevitably going to come from such a mission, the influence they have on the scope of human exploration to the red planet may influence where and for how long astronauts or robotic missions travel in the inner solar system. Resource utilization of NEOs and NEAs could support travel to Martian moons or newly discovered Martian Trojans, should they come.

Finally, as the dynamics of space weather are bound to play a factor in the execution of human exploration missions, the previously proposed space weather monitoring systems at various Lagrange points can communicate unanticipated changes to planners on Earth and Martian astronauts with enough warning to execute safety protocols to protect personnel and equipment.

Adaptability becomes not only a function of Systems Theory to study the application of Lagrange points supporting the human exploration of Mars, but an essential component in this complex and dynamic system-of-systems that will be required.

**Supporting the Hypotheses**

Early in this chapter we introduced the first of two hypotheses: Missions utilizing Lagrange points exhibit a high degree of systemic interdependence, influencing their design and operational strategies. This Systems Theory approach to understanding the application of Lagrange points in supporting the human exploration of Mars becomes apparent. Each component – space weather monitoring, communication, and the movement of astronauts from Earth to the surface of Mars and back – relies on the others for data, resources, and support, underlining a mesh of relationships vital for mission success. The dynamic interplay between various mission elements creates feedback loops that will influence the design and scope of future missions to Mars and possible other locations in the Martian vicinity. The adaptability of the utilization of Lagrange points supports changing operational strategies that could be employed as technological advancements regarding propulsion and spacecraft design. Each aspect of the system's design is crafted with the understanding that it must work within a complex web of relationships, adapting to feedback, and prepared for integration with future developments.

The second hypothesis: The integrated analysis of missions at Lagrange points, grounded in Systems Theory, reveals optimization pathways that are not apparent when missions are analyzed in isolation is directly supported by highlighting how viewing these missions through systemic, holistic lens can uncover optimization pathway that remain hidden when missions are considered in isolation. Interdependencies are revealed if space weather monitoring and communication relay systems share data or infrastructure, since there are possibilities of both being located at the same Lagrange location. Shared resources create efficiency where a relay satellite at Sun-Mars L4 point can serve multiple missions and can reduce the need for each mission to launch its own expensive communication hardware. Resiliency develops through feedback loops created from satellite constellations and launch and recovery platforms at the Sun-Mars L1 points. Non-linear interaction could develop as the adaptability of Lagrange points becomes better explored that may facilitate the movement of astronauts to Near Mars Objects (NeMOs) and Near Mars Asteroid (NeMAs). This systemic approach underlines the hypothesis by demonstrating an integrated analysis not only provides a comprehensive understanding of the complex dynamics as play among mission at Lagrange points, but also identifies opportunities for optimization that are not evident when considering missions individually.

System Theory offers a framework for viewing the ensemble of missions as a cohesive unit, where each component contributes to and benefits from the system, thereby revealing their interdependency and potential optimization pathways.

# CHAPTER FOUR: METHODOLOGY

**Chapter Introduction**

This chapter outlines the systematic methodology employed to explore the application of Lagrange points within various domains of space missions - astronomical research, space exploration, space resource utilization, and national defense (with space communications integrated throughout each area of focus). The focus of this study revolves around two primary hypotheses developed from the theoretical framework rooted in Systems Theory:

1. **H1:** Missions utilizing Lagrange points exhibit a high degree of systemic interdependence, influencing their design and operational strategies.
2. **H2:** The integrated analysis of missions at Lagrange points, grounded in Systems Theory, reveals optimization pathways that are not apparent when missions are analyzed in isolation.

To evaluate these hypotheses, a systematic review was conducted to aggregate and synthesize existing research on the specified areas of focus. This methodology was chosen due to its robustness in handling complex datasets and its effectiveness in uncovering underlying patterns and dependencies that align with Systems Theory. The systematic review aims not only to validate the interdependencies hypothesized but also to highlight efficiencies and innovations in mission design that emerge from a holistic analysis of systems involving Lagrange points.

By comprehensively exploring these aspects, this methodology supports the construction of a nuanced understanding of how Lagrange points function as pivotal elements within broader space mission architectures. This approach ensures that the study is grounded in rigorous academic standards while providing meaningful insights into the strategic utilization of Lagrange points.

The following sections detail the specific processes of the systematic review, including data sourcing, search strategy, selection criteria, and synthesis methods ensuring transparency and replicability of the research.

**Research Design**

This study adopts a qualitative research design, utilizing a systematic literature review and theoretical analysis to explore the application of Lagrange points within various space mission concepts. This approach is tailored to evaluate the interdependencies and potential optimizations in space mission designs involving Lagrange points, as hypothesized in the theoretical framework.

*Systematic Literature Review*

A comprehensive review of literature was conducted focusing on the use of Lagrange points in space missions across various domains such as astronomical research, space exploration, space resource utilization, and national defense. This review also includes an examination of the overarching theme of space communication. Key information regarding the integration and impact of Lagrange points in mission designs is extracted. This includes roles, dependencies, technological implications, and any noted efficiencies or challenges.

The review utilized academic search engines, including Google Scholar and the Richard D. Trefry university library, which often linked applicable articles and textbooks in academic journal publishers. Additionally, the use of academically-based Artificial Intelligence search systems such as Elicit and Litmaps were used to visualize the connected papers identified from the search in the previously mentioned databases. This allowed for the discovery of more relevant literature in the area of focus that was not listed in either Google Scholar or the university library, thereby expanding the base of information for which the analysis could be performed.

Search terms were carefully selected to capture the full spectrum of relevant literature by searching for the primary and secondary body in various Lagrangian systems. Combinations such as "Sun-Earth Lagrange point," "Sun-Earth L1", and "Sun-Earth Lagrangian System" were used for several popular Lagrangian system models in the Solar System to include the Sun, all major planets, and some moons. Papers were then sorted according to the area of focus previously discussed that they most closely aligned with. The digital object identifiers (doi) of published papers were fed into Litmaps to visualize what papers had been published after and referenced the seed paper and to understand how the research proposed in the original paper had (or if it had) evolved to reflect contemporary research philosophies or actual mission concepts.

Selection criteria whether to be included in or excluded from the literature review evaluated if the paper discussed the use of Lagrange points within the concept of operational space mission, theoretical models, or proposed applications. Initially, only papers published within the last 20 years were to be included to ensure relevancy; however, after an initial review, it was decided to not set a backward time limit on the published researched due to the required volume of literature required to make a substantial review possible. Some theoretical works which had publication dates dating back several decades were included in the background discussion section of the literature review for the purposes of establishing early concepts and theories that are discussed and evaluated more thoroughly is more recent works. Papers primarily centered on the mathematical or theoretical aspects of Lagrangian mechanics without direct application to space mission planning or implementation were excluded. While the foundational mathematics of Lagrange points is crucial for theoretical physics and advanced mathematics, this review specifically targeted practical applications in space mission. Papers discussing trajectory planning were included into the space resource utilization or space exploration areas of focus. Non-English publications were also excluded due to the language constraints of the author unless a translation of the publication was available in the publisher's database.

*Theoretical Application*

Each identified application of Lagrange points is analyzed through the lens of Systems Theory. This analysis emphasizes understanding the interconnections and feedback mechanisms within the space mission systems. A specific case study, Dr. Azita Valinia of NASA Jet Propulsion Laboratory's proposal for advanced reconnaissance missions utilizing Lagrange points ahead of the human exploration of Mars, is examined in depth. The proposal is scrutinized to identify how various mission elements, such as spacecraft deployment, data relay, and mission control, depend on functionalities provided by assets positioned at Lagrange points. This analysis reveals the crucial role these points play in ensuring mission flexibility and redundancy. Feedback loops within the mission design are identified and discussed. For instance, the proposal includes feedback from preliminary data gathered at Lagrange points, which influences subsequent decision on mission trajectory and focus. This feedback is vital for adaptive mission planning, allowing adjustment based on real-time environmental data and mission outcomes.

By applying Systems Theory, potential improvements in the mission design are suggested. These improvements include redundancy in communication pathways, better integration of mission control with automated systems at Lagrange points, and more robust data analysis capabilities that can dynamically alter mission parameters in response to incoming data streams. In this regard, Systems Theory not only aids in understanding the current state of mission designs but also serves as a prognostic tool. It helps predict how changes in one part of the system might have ripple or cascading effects through the entire system, enabling better preparation and robustness in mission planning.

*Synthesis and Interpretation*

Insights gathered from the literature review and theoretical analysis are synthesized to address the research hypotheses. This involves linking the conceptual underpinnings of Systems Theory with practical applications at Lagrange points discovered in the literature review.

The support for each hypothesis is assessed based on the synthesized insights. That section discussed the extent to which the literature substantiates the proposed interdependencies and optimization pathways, as well as how concepts from unrelated missions may support each other's objectives.

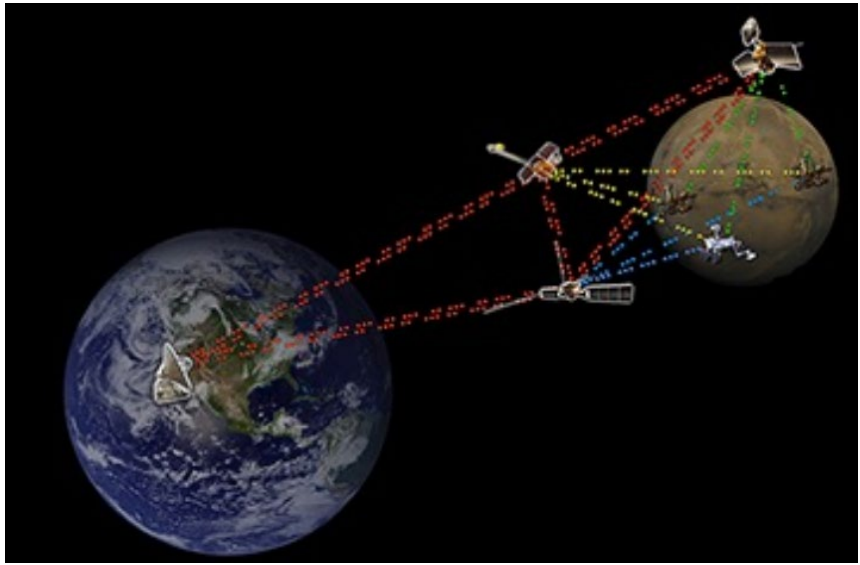

Figure 7 – Dr. Valinia's mission proposes a Lagrange point architecture for space weather and communication applications.

Source: *Valinia* (2023)

# CHAPTER FIVE: FINDINGS AND RESULTS

**Chapter Introduction**

This chapter presents the findings of the systematic review of Lagrange point applications in space missions, focusing on the four identified areas of focus: astronomical research, space exploration, space resource utilization, and national defense, along with the overarching theme of space communication. The results stem from a comprehensive analysis of the literature, filtered through the lens of Systems Theory to identify interdependencies and potential optimization pathways within and across the different applications.

The chapter begins by summarizing the principal finding related to each area of focus, highlighting the roles and dependencies of Lagrange points in the design and execution of space missions. The subsequent sections dive deeper into the synthesized insights, evaluating how these findings support the hypotheses posited in the theoretical framework chapter – that missions utilizing Lagrange points exhibit a high degree of systemic interdependence and that integrated analysis reveals optimization pathways that are not apparent when missions are analyzed in isolation.

**Astronomical Research**

In the realm of astronomical research, Lagrange points are primarily valued for their stable gravitational environment and unobstructed views of the observable universe or Solar System celestial objects. The review uncovered those missions stationed at these points, such as the James Webb Space Telescope at the Sun-Earth L2, the ISEE-3/ICE telescope at the Sun-Earth L1, and the STEREO Observatory at the Sun-Earth L4 and L5 positions, benefit significantly from reduced fuel requirements for orbit corrections, which in turn extends mission lifetimes and reduces operational costs. This finding underscores the first hypothesis, illustrating how the strategic positions at Lagrange points leverage the gravitational balance to optimize mission sustainability and effectiveness.

The literature supporting this area of focus was extensive, and despite the author's diligent efforts and the systematic nature of the review, it is inevitable that some proposal was unintentionally not included. A common subject in this area of focus was the Sun and the study of heliophysics and the space weather generated by it. This is likely, as uncovered in the research, that space weather heavily influences and affects all other mission types covered in the research. The very first application of Lagrange points (ISEE-3/ICE) made this apparent, and new proposals continue to be published making the Sun the subject of those proposals.

**Space Exploration**

The exploration of space, particularly through crewed and robotic missions to Mars and other celestial bodies, has demonstrated a multifaceted reliance on Lagrange points for communication relays and as staging areas for spacecraft. For instance, proposals for Mars missions have suggested using Sun-Mars L1 and L5 as critical waypoints for assembling and servicing mission components. This not only enhances the logistical feasibility of long-duration missions but also embeds a layer of adaptability and redundancy into the mission architecture.

Exploration missions utilizing the Sun-Moon Lagrangian system provide one of the strongest endorsements of the second hypothesis, where optimization pathways found through Systems Theory feedback loops influence missions to Mars that utilize those relevant Lagrange points. Space exploration missions utilizing Lagrange points have been frequently discussed in the literature, providing extensive and diverse applications, second only to astronomical research applications. The predominance of the literature reviewed discussed the applications for lunar and Martian exploration, indicating a collective focus across the field of research as those destinations become increasingly feasible with the technology required to deliver personnel and equipment to them advancing.

**Space Resource Utilization**

The efficient and innovative use of resources found within the Solar System to support a multitude of mission types is critical for the continued advancement of space and astronomical knowledge. The heavily utilized and highly valued gravitational stability of Lagrange points becomes the most prominent resource found in the literature. While unconventional in the understanding of a "resource", the ability to "park" a spacecraft or other object at any of the five Lagrange points for extended periods is a valuable application. As a resource, the stability of a Lagrange point is scarce (as in found in local environments and not widespread), cannot be artificially replicated, and offers significant utility in reducing mission costs and complexity.

The parking of other objects at Lagrange points for their extraction and utilization compounds this benefit. The literature identified several applications for moving asteroids, comets, and other objects to various Lagrange points for the extraction of water that will supplement and extend human missions to Mars or long-term habitation missions at L4 and L5 positions. Here, both hypotheses are supported, and as discussed in the case study in the theoretical framework chapter, new and innovative applications of Lagrange points for space resource utilization are uncovered when examining them as a component within the larger system and recognizing their interdependency on the entire space mission system.

**National Defense Applications**

This area of focus was revealed to be the thinnest in terms of published material to review, likely due to the sensitive nature of the application. Despite the constraint of only using public or unclassified material, the literature provided some insights and reviews that are not frequently covered in other discussions of Lagrange point applications. The use of the Earth-Moon Lagrangian system is considered more frequently than most other applications researched, due to the proximity of those points to Earth itself. The literature identified uses for communication strategies already in use by militaries that are the competitors to the United States. Kinetic strike capabilities, reviewed in the literature with a focus on Near Earth Objects that might threaten the planet, could easily be refocused on traditional military targets should efforts in policy and diplomacy fail.

Additionally identified were the use of Earth-Moon Lagrange points for trajectory planning within Cislunar space. This application supports the first hypothesis as it identifies that the Lagrange points themselves, when recognized as a node within a larger mission, become critical and dependent components in a complex and dynamic system and system of systems.

**Theoretical Framework Assessment**

The selection of Systems Theory as the foundational framework for this thesis was predicated on its robust ability to evaluate the complex interdependencies inherent in space missions utilizing Lagrange points. System Theory provides a proven lens to not only facilitate a deeper understanding of how different mission components interact, but also underscore the necessity of considering these missions as holistic systems rather than isolated operations.

Throughout the research, Systems Theory was crucial in the analysis of stated mission objectives from the literature and support the hypotheses that more optimal approaches, through the application of Lagrange points, could be identified and implemented. Systems Theory aided in illuminating the hypothesis of interconnectedness these missions sometimes identified or failed to identify. Most present during the case study, Systems Theory identified redundancy opportunities in space weather monitoring systems and space communication systems that had precedence with cited applications in the literature review. Feedback loops in lunar exploration missions then refine and improve on the scope and application of these systems for Martian exploration missions. The hypotheses were supported and reinforced through a comprehensive application of the theory, which is based on three principal pillars: the inherent interdependence of system components that work together to achieve a desired end state, the influence of feedback loops on system behavior, and the adaptability of system to their environment and inputs.

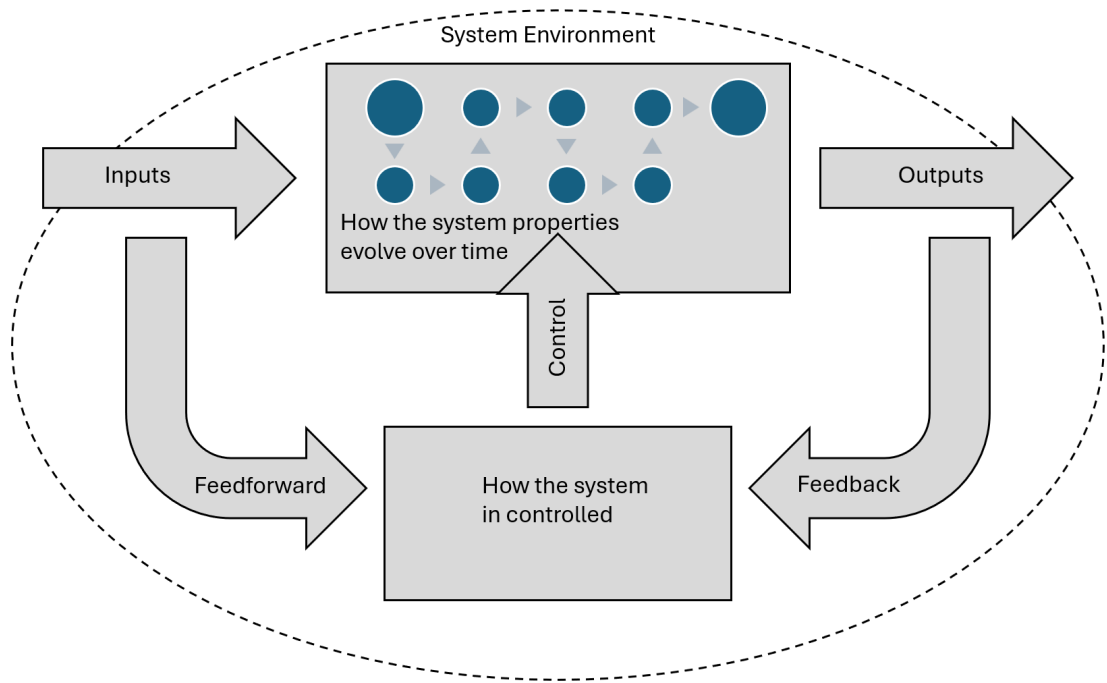

Figure 8 – Conceptual diagram of the principal components of Systems Theory as understood from Meadows (2008) and Skyttner (2002). Created by the author.

# CHAPTER SIX: DISCUSSION AND CONCLUSION

**Chapter Introduction**

This thesis embarked on a systematic exploration of Lagrange point applications within the realms of different types of space missions by employing a multifaceted approach grounded in Systems Theory. The comprehensive literature review revealed a diverse set of applications across astronomical research, space exploration, space resource utilization, and national defense, each underscored by the critical, yet often understated, role of space communications. These domains, interconnected through a complex web of technological, operational, and strategic dependencies, highlight the intricate nature of space missions and the pivotal role that Lagrange points play within them.

The application of Systems Theory proved instrumental in unraveling the complexities of these dependencies. By adopting this framework, the thesis was able to go beyond traditional analyses and uncover the nuanced interplay of mission components, specifically during the case study, and offer insights into systemic efficiencies and potentials enhancements. This not only validated the hypothesis that Lagrange point missions exhibit a high degree of systemic interdependence, but also supported the argument that a holistic, integrated analysis reveals optimization pathways otherwise obscured when missions are considered in isolation.

**Implications**

The findings of this research have significant implications for future space mission planning and execution. This thesis demonstrated the value of Systems Theory in analyzing the use of Lagrange points and contributed to a more nuanced understanding of space infrastructure planning and development. By forcing a holistic perspective to the mission architecture, we have exposed areas of opportunity for mission planners to exploit these naturally occurring gravity wells to host redundant platforms for monitoring and communication stations, flexible and adaptive locations for resource exploitation or long-term habitation, and system refinements and improvements built from feedback mechanisms that influence the future behavior, scope, and efficiency of related missions. When analyzed independently and separately, these insights can be lost as a mission planner's focus narrows to fit a limited objective with specific limitations. Systems Theory forces a dynamic criterion for problem solving that is better aligned with a dynamic mission.

**Applications**

The findings in this thesis may be of value for policymakers, scientists, and engineers involved in space mission planning and execution.

Policymakers could use the findings to prioritize investments in space infrastructure that leverage Lagrange points for strategic benefits. Understanding the interdependencies identified through Systems Theory, this could guide the funding of platforms that enhance resilience and multi-use efficiency for space exploration missions as identified in the case study. Additionally, the research highlights the importance of multiagency and indeed, international, cooperation when designing "Lagrange nodes" for dynamic mission enhancement. The research also identifies how easily current space resource applications for Earth-Moon Lagrange points could easily be militarized for national defense by the United States or its competitors.

For scientists in the field of astrophysics and space science, the insights on the gravitational stability of Lagrange points are likely already well understood. However, the detailed review of current applications and future proposals could serve as a reference when planning future research and even reserving observational time on existing platforms. A collection of established literature on the applications of Lagrange points for astronomical research enables researchers to explore novel uses of these positions and potentially highlights underused or unexplored applications as the review identified in the Sun-Neptune system or the Lagrangian systems of Jupiter, Saturn, Uranus, and Neptune and their respective moons and natural satellites.

For engineers and designers, the findings can be utilized to aid in the optimization for Lagrange points uses on spacecraft and space architecture. This could include energy-efficient trajectories from Earth to other celestial destinations through utilizations of the "parking spot" reference for Lagrange point gravitational stability. These points can extend planning factors by eliminating high fuel consumption rates for station keeping and allowing a craft in the Lagrange point to simply wait until that point better aligns with the desired object or launching point. When we used Systems Theory to identify the interdependency of each component of the space mission, it forced us to consider not just the primary mission objective of a specific platform, but how this component can support one or more other objectives, thereby enhancing mission success and reducing the costs associated with the development, launch, and maintenance of multiple unique systems "parked" at the same spot.

Practical applications now include a network of communication satellites at various Lagrange points in the inner Solar System that offer continuous, stable, and long-range communication for a multitude of space exploration missions to the Moon, Mars, and within Cislunar space. Redundant nodes for data relay within the inner Solar System mitigate the risk associated with communication failure from one or two dedicated platforms becoming inoperable. This network of nodes could support crewed and robotic missions with objectives in every area of focus as identified in the review of literature. Additionally, supply, fuel, and water stations at the Earth-Moon, Sun-Earth, and Sun-Mars Lagrange points can facilitate longer and more ambitious missions, reducing the need for frequent return missions to Earth and lowering the associated costs with launching specifically fuel and water from Earth to support missions to the Moon or Mars.

**Limitations and Future Research**

While the author attempted a thorough and systematic review of all current literature associated with Lagrange point applications for space missions, one fact became abundantly clear early in the research: the research and proposals for such uses is already extensive and continues to be published even at the time of the writing of this thesis. This indicates that this is an ever-evolving and changing area of research, where applications from as little as a few years ago may quickly become operational or outdated in their reasoning. It should therefore be understood that thesis reflects the most current and relevant theories that are presented at the time of its publication.

The research is also limited to the scope of practical applications and approaches to traditional mission objectives put forth by research institutions and government and private agencies for the utilization of Lagrange points to meet determined and measurable goals. As the literature review identified, a concept that is considered impractical at the time of publication may become completely practical with the unforeseen advancement in spacecraft or fuel source technology that brings more audacious objectives into focus. Those applications now are limited to what research has been peer reviewed and published in textbooks and journal articles available online and occasionally through policy and planning institutes and institutions. On-going and unfinished research that attempts to exploit the use of Lagrange points for entirely practical and reasonable applications could not be reviewed.

The existing literature is heavily focused on the inner Solar System and the astronomical applications for the use of the Sun-Earth Lagrange system, except for the Sun-Earth L3 point, which lies 180 degrees ahead of or behind Earth along its orbital path. Aside from heliophysics, it remains uncertain what practical applications for this L3 point, or any L3 point in the Solar System there could be. For much of the research, either the main or secondary body in the Lagrangian system was the identified subject of research or exploration mission that utilized the Lagrange point in question. The L3 point, being far removed from the secondary body in any Lagrangian system provides no direct application for the study of or use of the secondary body. The distance the L3 point is away from the primary body dramatically limits the information that can be gathered or collected from it, with the Sun being the primary exception to that rule. It remains uncertain what advantage the Earth-Moon L3 or the Sun-Mars L3 point provides that could not be adequately covered by another collinear point in that system.

The outer Solar System remains ripe for new proposals and potential uses, especially when considering the efficiency that could be enhanced when planning trajectories through Lagrange points to outer destinations. Even then, the fuel required for transporting a spacecraft, station, or crew from Earth to the gas giants will be significant, and settling in Lagrange points provides an area where minimal station-keeping maneuvers are required, thus reducing the amount of fuel carried from Earth. Harvesting fuel and water from objects at Lagrange points between Earth and the desired outer Solar System destination further extends that mission objective and reduces the material required to bring from Earth. Astronomical research could also be enhanced by studying those objects already at outer Solar System Lagrange points or even objected moved into those points. NASA's Lucy mission is a visible mission and may inspire missions to the Sun-Neptune Trojans at the L4 and L5 positions.

Farquhar, R. W. (2012). Lunar communications with libration-point satellites. *Journal of Spacecraft and Rockets*, *4*(10), 1383–1384. https://doi.org/10.2514/3.29095

Fraser, C. (1983). J. L. Lagrange's early contributions to the principles and methods of mechanics. *Archive for History of Exact Sciences, 28*(3) 197-241. https://www.jstor.org/stable/41133689

Frazier, S. (2020). 4,000th Comet discovered by ESA & NASA Solar Observatory. NASA.gov. https://www.nasa.gov/solar-system/4000th-comet-discovered-by-esa-nasa-solar-observatory/

Gruenwald, J. (2014). Human outposts on Mars: Engineering and scientific lessons learned from history. *CEAS Space Journal*, *6*(2), 73–77. https://doi.org/10.1007/S12567-014-0059-8/METRICS

Hall, R. C. (2007). The evolution of U.S. national security space policy and its legal foundations in the 20th century. *Journal of Space Law*, *33*. https://heinonline.org/HOL/Page?handle=hein. journals/jrlsl33&id=197&div=&collection

Hapgood, M. (2017). L1L5Together: Report of workshop on future missions to monitor space weather on the sun and in the solar wind using both the L1 and L5 Lagrange points as valuable viewpoints. *Space Weather*, *15*(5), 654–657. https://doi.org/10.1002/2017SW001652

Horner, J., & Lykawka, P. S. (2010). The Neptune Trojans-a new source for the centaurs? *Monthly Notices of the Royal Astronomical Soc*iety *402*, 13–20. https://doi.org/10.1111/j.1365-2966.2009. 15702.x

Howard, R. L., & Seibert, M. (2010). *Lagrange-based options for relay satellites to eliminate earth-mars communications outages during solar superior conjunctions*. National

Aeronautics and Space Administration. https://ntrs.nasa.gov/api/citations/20205007788/downloads/FINAL%20-%20Lagrange-Based%20Options%20for%20Relay%20Satellites%20to%20Eliminate%20Earth-Mars%20Communications%20Outages%20During%20Solar%20Superior%20Conjunctions.pdf

Howel, K.C., Barden, B.T., & Lo, M.W. (2020). Application of dynamical systems theory to trajectory design for a libration point mission. *Journal of Astronautical Sciences, 45,* 161-178. https://doi.org/10.1007/BF03546374

Iliopoulos, N., & Esteban, M. (2020). Sustainable space exploration and its relevance to the privatization of space ventures. *Acta Astronautica*, *167*, 85–92. https://doi.org/10.1016/J.ACTAASTRO.2019.09.037

John, K. K., Graham, L. D., & Abell, P. A. (2015). *Investigating trojan asteroids at the L4/L5 sun-earth Lagrange points.* NASA Johnson Space Center. https://ntrs.nasa.gov/api/citations /20150002915/downloads/20150002915.pdf

Kaplan, S. (2020). *Eyes on the prize the strategic implications of cislunar space and the moon* (CSIS Reports). Center for Strategic & International Studies. https://aerospace.csis.org/wp-content/uploads/2020/07/20200714_Kaplan_Cislunar_FINAL.pdf

Lagrangian Point, (2024). *Toppr*. https://www.toppr.com/guides/physics/gravitation/lagrangian-point/

Levison, H. F., Olkin, C. B., Noll, K. S., Marchi, S., Bell, J. F., Bierhaus, E., Binzel, R., Bottke, W., Britt, D., Brown, M., Buie, M., Christensen, P., Emery, J., Grundy, W., Hamilton, V. E., Howett, C., Mottola, S., Pätzold, M., Reuter, D., ...Wong, I. (2021). Lucy mission to the trojan asteroids: Science goals. *The Planetary Science Journal*, *2*(5), 171. https://doi.org/10.3847/PSJ/ABF840

Lin, H., Chen, Y., Holman, M., Ip, W., Payne, M., Lacerda, P., Fraser, W., Gerdes, D., Bieryla,54